\documentclass [12pt]{article}

\usepackage{amsmath}
\usepackage{amsthm}

\newcommand{\keywords}[1]{\par\strut\vskip0.3cm\parindent 0pt
	\footnotesize{ KEYWORDS: #1}}
\newcommand{\institute}[1]{\\{\small #1}}
\newcommand{\address}[1]{\\{\small #1}}
\newcommand{\email}[1]{\\{\small Email: #1}}

\newtheorem {lemma}{Lemma}
\newtheorem {proposition}{Proposition}
\newtheorem {definition}{Definition}
\newtheorem {theorem}{Theorem}

\newenvironment {acknowledgement}
  {\section*{Acknowledgements}}
  {}

\def\Io{{\bf I}}  
\def\Ro{{\bf R}}  
\def\Co{{\bf C}}  

\begin{document}

\title{Covariance systems}

\author{
  Jan Naudts and 
  Maciej Kuna\footnote{On leave from the Technical University of Gdansk}
  \institute{
  Departement Natuurkunde, Universiteit Antwerpen UIA
  }
  \address{
  Universiteitsplein 1, B-2610 Antwerpen, Belgium
  }
  \email{
  naudts@uia.ua.ac.be, kuna@uia.ua.ac.be
  }
} 

\date{v6, September 2000}

\maketitle

\begin{abstract}

We introduce new definitions of states and of representations of covariance 
systems. The GNS-construction is generalized to this context. It associates a 
representation with each state of the covariance system. Next, states are 
extended to states of an appropriate covariance algebra. Two applications are 
given. We describe a nonrelativistic quantum particle, and we give a simple 
description of the quantum spacetime model introduced by Doplicher et
al \cite{DFR94,DFR95}.

\end{abstract}

\subsection* {Introduction}

Consider an action $\sigma$ of a group $X$ as automorphisms
of a $C^*$-algebra $\cal A$. A $*$-representation $\pi$ of $\cal A$
is {\sl $X$-covariant} if a unitary representation $U$ of $X$ exists
such that
\begin{equation}
\pi(\sigma_xa)=U(x)\pi(a)U(x)^*
\label{covariant}
\end{equation}
holds for all $a\in{\cal A}$
and $x\in X$.
A state $\omega$ of $\cal A$ is $X$-covariant if its GNS-representation is
$X$-covariant.
Doplicher et al \cite {DKR66} proved a one-to-one relation
between covariant $*$-representations of $\cal A$ and
$*$-representations of the crossed product algebra ${\cal A}\times_\sigma X$.
In quantum mechanics also projective representations of the
relevant symmetry group $X$ are important. A $*$-representation $\pi$
of $\cal A$, which is only covariant in the sense that (\ref{covariant})
holds w.r.t.~a projective representation $U$ of $X$, cannot extend to
a $*$-representation of ${\cal A}\times_\sigma X$
because then $U$ would not be projective. This leads to the situation
that the covariance algebra ${\cal A}\times_\sigma X$ has to be
replaced by some other algebra which depends in general on the choice of $\pi$.
An elegant formalism to deal with this situation is presented here.
It introduces new concepts of representation and state of a covariance system.
The GNS-construction, well-known for states of $C^*$-algebras,
is generalized and associates a representation with each state of
the covariance system. Next, it is shown that each state of the
covariance system extends to a state of an appropriate covariance algebra.

Our results on covariance systems combine well with a new $C^*$-algebraic 
approach to quantum mechanics and quantum field theory. The quantum mechanics of 
a single nonrelativistic particle is described by the covariance system 
consisting of the group of spatial shifts $\Ro^3,+,$ acting on an abelian 
algebra of functions of position. If projective representations are allowed then 
the group of shifts can be replaced by the full Galilei group. Indeed,
Levy-Leblond \cite{LLB63} has shown, using results of Bargmann \cite{BV54},
that the physically 
relevant representations of the Galilei group are projective representations, 
labeled by a free parameter, which is proportional to the mass of the particle. 
In addition, projective representations of the subgroup of rotations describe 
the spin of the particle. One concludes that, in the terminology of the present 
paper, each nonrelativistic particle is described by a state of the covariance 
system. It has a mass, and can have a spin.

As a second example we consider the model of quantum spacetime introduced by 
Doplicher et al \cite{DFR94,DFR95}.
We restrict ourselves to the description of a 
single particle. Its configuration space $\Sigma$ is a 4-dimensional manifold in
$\Ro^6$. Shifts in Minkowski space and in momentum space act on $\Sigma$ 
in the trivial way. An explicit expression is given for a quasifree state of the 
resulting covariance system. It involves a $C^*$-multiplier with values in the 
algebra of continuous complex functions of $\Sigma$. It has a nontrivial GNS-
representation in which the shift groups have projective representations. Their 
generators are position and momentum operators satisfying noncanonical 
commutation relations.

A third example, the electromagnetic radiation field, will be discussed elsewhere.

\section{Covariance systems}

\subsection{$C^*$-multipliers}

The following definition introduces a pair
of objects $(\xi,\sigma)$, where $\xi$ generalizes the concept of
multiplier to maps with values in a $C^*$-algebra,
and where $\sigma$ is a twisted representation of a group as
automorphisms of a $C^*$-algebra.
Operator valued multipliers have been studied in \cite{BS70,NJ99}.

\begin{definition}
A (left) $C^*$-multiplier of a locally compact group $X$ with associated
twisted representation $\sigma$ is
a measurable map $\xi$ of $X\times X$ into the unitary elements
of the multiplier algebra $M({\cal A})$ of a $C^*$-algebra $\cal A$ satisfying
\begin{equation}
\xi(x,e)=\xi(e,y)=\Io,
\qquad x,y\in X,
\end{equation}
together with a map $\sigma$ of $X$ into the automorphisms of $\cal A$
such that $\sigma_e$ is the identity transformation and
\begin{equation}
\sigma_x\xi(y,z)=\xi(x,y)\xi(xy,z)\xi(x,yz)^*,
\qquad x,y,z\in X
\label{modcocycleprop}
\end{equation}
and
\begin {equation}
\sigma_x\sigma_ya=\xi(x,y)(\sigma_{xy}a)\xi(x,y)^*,
\qquad x,y\in X,a\in {\cal A}
\label {projtrans}
\end {equation}
\end{definition}

Throughout the paper we assume that $\xi$ is
continuous in a neighborhood of the neutral element $e$ of $X$.
If $\xi(x,y)$ is a multiple of $\Io$
for all $x,y\in X$ then $\xi$ is also called a {\sl cocycle}.

A right $C^*$-multiplier $\zeta$ satisfies
\begin{equation}
\sigma_y^{-1}\zeta(z,x)=\zeta(zx,y)^*\zeta(z,xy)\zeta(x,y),
\qquad x,y,z\in X
\label{rightmult}
\end{equation}
instead of (\ref{modcocycleprop})
and
\begin {equation}
\sigma_x\sigma_ya=\sigma_{xy}(\zeta(x,y)a\zeta(x,y)^*),
\qquad x,y\in X,a\in {\cal A}
\end {equation}
instead of (\ref{projtrans}).

\subsection{Covariant states}

The following definition can be found in the literature, see e.g.~\cite{TM79}.
\begin{definition}
\label{covsys}
A {\rm covariance system} is a triple $({\cal A},X,\sigma)$ consisting of
a $C^*$-algebra $\cal A$, a locally compact symmetry group $X$,
and a continuous action $\sigma$ of $X$ as automorphisms of $\cal A$.
\end{definition}

The definition generalizes that of a dynamical system as often found in the 
literature (take $X=\Ro,+$, and $t\in\Ro\rightarrow \sigma_t$ the time evolution 
of the quantum system). Here we have rather different applications in mind. 
Examples are given further on.

The next definition is new.

\begin{definition}
\label{covstate}
A {\rm state} of the covariance system $({\cal A},X,\sigma)$
is a measurable map $\omega$
of $X\times X$ into the continuous linear complex valued functions of $\cal A$,
having the following properties.
\begin{itemize}
\item
(positivity)
For all $n>0$ and for all possible choices of
$\lambda_1,\ldots,\lambda_n$ in $\Co$,
of $x_1,\ldots,x_n$ in $X$,
and of $a_1,\ldots a_n$ in $\cal A$ is
\begin{equation}
\sum_{j,k=1}^n\lambda_j\overline{\lambda_k}\omega_{x_j,x_k}(a_k^*a_j)
\ge 0
\end{equation}
\item
(normalization)
$\omega_{e,e}$ is a state of $\cal A$.
\item
(covariance)
There exists a right $C^*$-multiplier $\zeta$ of $X$
with values in the multiplier algebra $M({\cal A})$ of $\cal A$ such that
\begin{equation}
\omega_{x,y}(\sigma_za)=
\omega_{xz,yz}(\zeta(y,z)a\zeta(x,z)^*)
\label{zetadef}
\end{equation}
 for all $x,y,z\in X$ and $a\in {\cal A}$.
\item
(continuity)
For any $a\in{\cal A}$ the map $x,y\rightarrow \omega_{x,y}(a)$
is continuous in a neighborhood of the neutral element of $X$.
\end{itemize}
The state $\omega$ is {\rm faithful} if $\omega_{e,e}$
is faithful.
\end{definition}

Any $X$-covariant state $\omega$ of $\cal A$ defines a state (again denoted
$\omega$) of the covariance system $({\cal A},X,\sigma)$
by
\begin{equation}
\omega_{x,y}(a)=(\pi(a)U(x)^*\Omega,U(y)^*\Omega),
\qquad x,y\in X,a\in {\cal A}
\label{omegadef}
\end{equation}
where $({\cal H},\pi,\Omega)$ is the GNS-representation
of $\cal A$ induced by $\omega$, and $U$ is the covariant representation
of $X$. The converse is also true. If $\omega$ is a state of
$({\cal A},X,\sigma)$ then $\omega_{e,e}$ is an $X$-covariant
state of $\cal A$, provided we weaken the definition of
$X$-covariance to allow for projective representations.
This will be shown in section \ref{GNS}.

The following results are needed for technical reasons.
The next lemma generalizes the lemma of Schwarz.

\begin{lemma} Let $\omega$ be a state of $({\cal A},X,\sigma)$. Then
\begin{equation}
\left|
\sum_{j,k=1}^n\lambda_j\overline{\lambda_k}
\omega_{x_j,x_k}(a_k^*b_j - b_k^*a_j)
\right|^2
\le 4
\sum_{j,k=1}^n\lambda_j\overline{\lambda_k}\omega_{x_j,x_k}(a_k^*a_j)
\sum_{j,k=1}^n\lambda_j\overline{\lambda_k}\omega_{x_j,x_k}(b_k^*b_j)
\end{equation}
for all choices of $\lambda_1,\ldots,\lambda_n$ in $\Co$,
of $x_1,\ldots,x_n$ in $X$,
and of $a_1,\ldots a_n$ and $b_1,\ldots,b_n$ in $\cal A$.
\end{lemma}
The proof of the lemma is straightforward.
The lemma is now used to prove uniform continuity of
$\omega_{x,y}$.

\begin{proposition}
One has
\begin{equation}
\sum_{j,k=1}^n\lambda_j\overline{\lambda_k}\omega_{x_j,x_k}(a_k^*a_j)
\le \left(\sum_{j=1}^n|\lambda_j|\,||a_j||\right)^2
\end{equation}
for all choices of $\lambda_1,\ldots,\lambda_n$ in $\Co$,
of $x_1,\ldots,x_n$ in $X$,
and of $a_1,\ldots a_n$ in $\cal A$.
\end{proposition}

\begin{proof}
The statement is clearly true for $n=1$. Assume it to hold up to $n-1$.
One calculates, using the lemma with $b_j=i\delta_{k,n}a_n$,
\begin{eqnarray}
\sum_{j,k=1}^n\lambda_j\overline{\lambda_k}\omega_{x_j,x_k}(a_k^*a_j)
&\le&
\left(\sum_{j=1}^{n-1}|\lambda_j|\,||a_j||\right)^2
+|\lambda_n|^2\,||a_n||^2\cr
&+&
\sum_{j=1}^{n-1}\lambda_j\overline{\lambda_n}\omega_{x_j,x_n}(a_n^*a_j)
+
\sum_{j=1}^{n-1}\lambda_n\overline{\lambda_j}\omega_{x_n,x_j}(a_j^*a_n)\cr
&\le&\left(\sum_{j=1}^n|\lambda_j|\,||a_j||\right)^2
\end{eqnarray}
Hence the proof follows by induction.
\end{proof}

\subsection{Representations of a covariance system}

The following two definitions are obvious.

\begin{definition}
A representation of the covariance system $({\cal A},X,\sigma)$
is a triple $({\cal H},\pi,U)$ which consists of
a *-representation $\pi$ of $\cal A$ in a Hilbert space $\cal H$, and
a measurable map $x\in X\rightarrow U(x)$ into the unitary
operators of $\cal H$, with the properties that (\ref{covariant}) holds,
and that each normalized element $\psi$
of $\cal H$ defines an $X$-covariant state $\omega$ by (\ref{omegadef}).
An element $\psi\in{\cal H}$ is {\sl cyclic}
for the representation if the subspace spanned by
\begin{equation}
\{\pi(a)U(x)^*\psi:\,a\in{\cal A},x\in X\}
\end{equation}
is dense in $\cal H$.
\end{definition}

\begin{definition}
Two representations $({\cal H},\pi,U)$ and $({\cal H}',\pi',U')$
of $({\cal A},X,\sigma)$ are {\rm equivalent} if there
exists an isomorphism $V$ of $\cal H$ onto ${\cal H}'$
intertwining $\pi$ and $\pi'$, resp. $U$ and $U'$.
\end{definition}

\subsection{Projective representations of $X$}

\begin{proposition}
\label{projprop}
Let $\pi$ be a *-representation of $\cal A$
in a Hilbert space $\cal H$ and $x\rightarrow U(x)$
is a projective representation of $X$ in $\cal H$
in the sense that $U(e)=\Io$, $U(x)$ is unitary for all $x\in X$,
and a $C^*$-multiplier $\xi$ exists
with values in the center of $M({\cal A})$ such that
\begin{equation}
U(x)U(y)=\pi(\xi(x,y))U(xy),
\qquad x,y\in X
\label{projrepr}
\end{equation}
Assume that the twisted representation associated with $\xi$
coincides with the action $\sigma$ of $({\cal A},X,\sigma)$,
that $x\rightarrow U(x)$ is strongly continuous in a neighborhood of
the neutral element of $X$, and
that the covariance condition (\ref{covariant}) is satisfied.
Then $({\cal H},\pi,U)$ is a representation of the
covariance system $({\cal A},X,\sigma)$.
\end{proposition}

\begin{proof}
Each normalized element $\psi$ of $\cal H$
defines a covariant state $\omega$ by (\ref{omegadef}).
In particular, one has
\begin{equation}
U(xy)^*U(x)U(y)=\pi(\zeta(x,y))
\end{equation}
with
\begin{equation}
\zeta(x,y)=\sigma_{xy}^{-1}\xi(x,y)
\label{xizeta}
\end{equation}
Using this result covariance follows from
\begin{eqnarray}
\omega_{x,y}(\sigma_za)
&=&(\pi(\sigma_za)U(x)^*\psi,U(y)^*\psi)\cr
&=&(\pi(a)U(z)^*U(x)^*\psi,U(z)^*U(y)^*\psi)\cr
&=&(\pi(a)(U(xz)\pi(\zeta(x,z)))^*\psi,(U(yz)\pi(\zeta(y,z)))^*\psi)\cr
&=&\omega_{xz,yz}(\zeta(y,z)a\zeta(x,z)^*)
\end{eqnarray}

\end{proof}

The requirement that $\xi(x,y)$ commutes with all elements of $\cal A$ is needed 
because the action $\sigma$ of $X$ on $\cal A$ is not twisted with $\xi$ but is 
a representation of $X$ as automorphisms of $\cal A$.

\subsection{GNS-construction}
\label{GNS}

The GNS-construction for states of a $C^*$-algebra
can be generalized as follows.

\begin{theorem}
Let $\omega$ be a covariant state of the covariance system $({\cal A},X,\sigma)$.
There exists a representation $({\cal H},\pi,U)$ of
$({\cal A},X,\sigma)$ satisfying
\begin{itemize}
\item
$U$ is a projective representation of $X$
with multiplier $\xi$.
\item
(\ref{zetadef}) holds with $\zeta$ related to $\xi$ by (\ref{xizeta}).
\item
There exists a cyclic vector $\Omega$ of $\cal H$ such that
(\ref{omegadef}) holds.
\end{itemize}
The quadruple $({\cal H},\pi,U,\Omega)$ is unique up to equivalence
of representations, i.e.~ if $({\cal H}',\pi',U')$
is a representation and $\Omega'$ is a cyclic vector of ${\cal H}'$
satisfying (\ref{omegadef}) with $\psi=\Omega'$ and $\pi$ and $U$
replaced by $\pi'$ and $U'$, then there exists an isomorphism $V$
of $\cal H$ onto ${\cal H}'$ intertwining $\pi$ and $\pi'$,
resp.~$U$ and $U'$, and mapping $\Omega$ onto $\Omega'$.
\end{theorem}

\begin{proof}
Let ${\cal C}_c(X,{\cal A})$ denote the linear space of continuous
functions with compact support in $X$ and with values in $\cal A$.
A sesquilinear form is defined by
\begin{equation}
(f,g)=\int_X\hbox{ d}x\Delta(x)^{-1}\int_X\hbox{ d}y\Delta(y)^{-1}\,\,
\omega_{x,y}(g(y)^*f(x))
\label{scaldef}
\end{equation}
for all $f,g\in {\cal C}_c(X,{\cal A})$
($\Delta$ is the modular function of $X$).
From the positivity of $\omega$ follows that $(\cdot,\cdot)$
is a positive form.
Let us assume for simplicity of notations that $(\cdot,\cdot)$ is not degenerated
(it is easy to see that further definitions do not depend on the
choice of the representative of the equivalent class given by the kernel
of $(\cdot,\cdot)$). Then it is an 
inner product making ${\cal C}_c(X,{\cal A})$ into a pre-Hilbert space.
Let $\cal H$ denote its completion.

Define $\pi$ by
\begin{equation}
\pi(a)f(x)=af(x),
\qquad a\in {\cal A},f\in {\cal C}_c(X,{\cal A}),x\in X
\end{equation}
By linearity $\pi(a)$ extends to a linear operator with domain ${\cal C}_c(X,{\cal A})$.
If $a\ge 0$ then
\begin{eqnarray}
(\pi(a)f,f)
&=&\int_X\hbox{ d}x\Delta(x)^{-1}\int_X\hbox{ d}y\Delta(y)^{-1}\,
\omega_{x,y}(f(y)^*af(x))
\ge 0
\end{eqnarray}
This implies that $\pi(a)$ is bounded.
Since each element of $\cal A$ is a linear combination
of positive elements one concludes that
$\pi(a)$ is bounded for all $a\in{\cal A}$.

It is obvious that $\pi(a)\pi(b)=\pi(ab)$, that $\pi$ is linear,
and that $\pi(b)^*=\pi(b^*)$.
Hence $\pi$ is a *-representation of $\cal A$ in $\cal H$.

Next define a linear operator $U$ by
\begin{equation}
U(x)f(y)=\sigma_{x}\left[f(yx)\zeta(y,x)\right]
\qquad x,y\in X,f\in {\cal C}_c(X,{\cal A})
\label{udef}
\end{equation}
Note that one has $U(e)=\Io$.
A straightforward calculation gives
\begin{equation}
U(x)^*f(y)=(\sigma_x^{-1}f(yx^{-1}))\zeta(yx^{-1},x)^*
\label{auform}
\end{equation}
This expression can be used to verify that $U(x)$ is unitary.

Let $\xi$ be defined by (\ref{xizeta}).
Then a short calculation using (\ref{rightmult}) gives
\begin{eqnarray}
U(x)U(y)f(z)
&=&\sigma_{x}\left[\sigma_y\left[f(zxy)\zeta(zx,y)\right]\zeta(z,x)\right]\cr
&=&\sigma_{xy}\left[f(zxy)\zeta(x,y)\zeta(z,xy)\right]\cr
&=&\pi(\xi(x,y))U(xy)f(z)
\end{eqnarray}
which is (\ref{projrepr}).

Let $(u_\alpha)_\alpha$ be an approximate unit of $\cal A$.
For each neighborhood $v$ of $e$ in $X$ let $\delta_v$
be a positive function vanishing outside $v$ and satisfying
\begin{equation}
\int_X\hbox{ d}x\,\delta_v(x)=1
\end{equation}
Then the functions $(\delta_vu_\alpha)_{v,\alpha}$
form a Cauchy sequence in $\cal H$, converging to some
element $\Omega$.
From (\ref{auform}) follows now that
\begin{equation}
\int_X\hbox{ d}x\,\Delta(x)^{-1}\pi(f(x))U(x)^*\Omega=f
\label{frep}
\end{equation}
for all $f\in{\cal C}_c(X,{\cal A})$.
The latter implies
\begin{equation}
(f,g)=\int_X\hbox{ d}x\Delta(x)^{-1}\int_X\hbox{ d}y\Delta(y)^{-1}\,
(\pi(g(y)^*f(x))U(x)^*\Omega,U(y)^*\Omega)
\end{equation}
Comparison with (\ref{scaldef}) shows that
(\ref{omegadef}) holds.
From (\ref{udef}) and (\ref{auform}) follows immediately
that (\ref{covariant}) holds.
Hence, proposition \ref{projprop} asserts that
$({\cal H},\pi,U)$ is a representation of the covariance system.

Finally, we prove uniqueness up to equivalence of representations.
Define $V$ with domain ${\cal C}_c(X,{\cal A})$ by
\begin{equation}
Vf=\int_X\hbox{ d}x\,\Delta(x)^{-1}\pi'(f(x))U'(x)^*\Omega'
\end{equation}
It is straightforward to verify that $V$
extends to an isometry of $\cal H$ into ${\cal H}'$.
Because $\Omega'$ is cyclic $V$ is an isomorphism.

That $V$ intertwines $\pi$ and $\pi'$ is obvious. For $U$ and $U'$ one has
\begin{eqnarray}
U'(x)Vf
&=&U'(x)\int_X\hbox{ d}y\,\Delta(y)^{-1}
\pi'(f(y))U'(y)^*\Omega'\cr
&=&U'(x)\int_X\hbox{ d}y\,\Delta(y)^{-1}
\pi'(f(yx))U'(yx)^*\Omega'\cr
&=&U'(x)\int_X\hbox{ d}y\,\Delta(y)^{-1}
\pi'(f(yx)\zeta(y,x))U'(x)^*U'(y)^*\Omega'\cr
&=&\int_X\hbox{ d}y\,\Delta(y)^{-1}
\pi'\left(\sigma_x\left[f(yx)\zeta(y,x)\right]\right)
U'(y)^*\Omega'\cr
&=&VU(x)f
\end{eqnarray}
Finally, it is obvious that $V\Omega=\Omega'$.

\end{proof}

\subsection{Crossed product algebras}

A reader not interested in crossed product algebras may skip this section.
It is not needed for the sequel of the paper but is added to clarify the
relation between the present work and previous work on crossed product
algebras. Details found in \cite{NJ99} are not repeated here.

Let be given a state $\omega$ of a covariance system
$({\cal A},X,\sigma)$. Let $\zeta$ be the right multiplier associated with
$\omega$. Let $\xi$ be the left multiplier derived from $\zeta$ by (\ref{xizeta}).
The representation $\sigma$ together with $\xi$
determine a crossed product algebra ${\cal A}\times_\xi X$.
It is constructed as follows.
Let ${\cal L}_1(X,{\cal A})$ denote the linear space of
integrable functions of $X$ with values in ${\cal A}$.
A product law for elements of ${\cal L}_1(X,{\cal A})$
is given by
\begin{equation}
(f\times g)(x)=\int_X\hbox{ d}y\,f(y)\xi(y,y^{-1}x)\sigma_yg(y^{-1}x)
\end{equation}
An involution is given by
\begin{equation}
f^\star(x)=\Delta(x)^{-1}\xi(x,x^{-1})^*\sigma_xf(x^{-1})^*
\end{equation}
In this way ${\cal L}_1(X,{\cal A})$ becomes an involutive
algebra. By closure in an appropriate norm it becomes
the $C^*$-algebra ${\cal A}\times_\xi X$.

A linear functional $\overline\omega$ of ${\cal L}_1(X,{\cal A})$
is defined by
\begin{equation}
\overline\omega(f)=\int_X\hbox{ d}x\,\Delta(x)^{-1}
\omega_{x,e}(\xi(x^{-1},x)f(x^{-1}))
\end{equation}

\begin{proposition}
$\overline\omega$ extends to a state of ${\cal A}\times_\xi X$.
\end{proposition}

\begin{proof}

Positivity is verified as follows.
\begin{eqnarray}
& &{\overline\omega}(f^\star\times f)\cr
&=&\int_X\hbox{ d}x\,\Delta(x)^{-1}
\omega_{x,e}(\xi(x^{-1},x)(f^\star\times f)(x^{-1}))\cr
&=&\int_X\hbox{ d}x\,\Delta(x)^{-1}\int_X\hbox{ d}y\,\Delta(y)^{-1}\cr
&\times&
\omega_{x,e}(\xi(x^{-1},x)\xi(y,y^{-1})^*(\sigma_yf(y^{-1}))^*
\xi(y,y^{-1}x^{-1})\sigma_yf(y^{-1}x^{-1}))\cr
&=&\int_X\hbox{ d}x\,\Delta(x)^{-1}\int_X\hbox{ d}y\,\Delta(y)^{-1}\cr
&\times&
\omega_{x,e}(\sigma_y\left[
\xi(y^{-1}x^{-1},x)f(y^{-1}))^*f(y^{-1}x^{-1})
\right])\cr
&=&\int_X\hbox{ d}x\,\Delta(x)^{-1}\int_X\hbox{ d}y\,\Delta(y)^{-1}\cr
&\times&
\omega_{xy,y}(
\xi(y^{-1}x^{-1},x)\zeta(x,y)^*f(y^{-1}))^*f(y^{-1}x^{-1})
)\cr
&=&\int_X\hbox{ d}x\,\Delta(x)^{-1}\int_X\hbox{ d}y\,\Delta(y)^{-1}\cr
&\times&
\omega_{xy,y}(
\xi(y^{-1},y)\xi(y^{-1}x^{-1},xy)^*f(y^{-1}))^*f(y^{-1}x^{-1})
)\cr
&=&\int_X\hbox{ d}x\,\Delta(x)^{-1}\int_X\hbox{ d}y\,\Delta(y)^{-1}
\omega_{x,y}(
\xi(y^{-1},y)\xi(x^{-1},x)^*f(y^{-1}))^*f(x^{-1})
)\cr
&\ge&0
\end{eqnarray}

For each neighborhood $v$ of the neutral element of $X$
let $\delta_v$ be a positive normalized function with
support in $v$. Then $(\delta_v u_\alpha)_{v,\alpha}$
is an approximate unit of ${\cal L}_1(X,{\cal A})$.
One has
\begin{eqnarray}
\overline\omega(\delta_v u_\alpha)
&=&\int_X\hbox{ d}x\,\Delta(x)^{-1}\delta_v(x^{-1})\omega_{x,e}
(\xi(x^{-1},x)u_\alpha)
\end{eqnarray}
The latter tends to 1 because of continuity of 
$x\rightarrow \omega_{x,e}$ and $x\rightarrow\xi(x^{-1},x)$
in the vicinity of $e$.

One concludes that $\overline\omega$ is a positive normalized linear
functional on the involutive algebra ${\cal L}_1(X,{\cal A})$.
By continuity it extends to a state of ${\cal A}\times_\xi X$.

\end{proof}

Now let $({\cal H},\pi,U,\Omega)$ be the GNS-representation
of $({\cal A},X,\sigma)$ induced by $\omega$.
Then one has
\begin{eqnarray}
\overline\omega(f)
&=&\int_X\hbox{ d}x\,\Delta(x)^{-1}\omega_{x,e}(\xi(x^{-1},x)f(x^{-1}))\cr
&=&\int_X\hbox{ d}x\,\Delta(x)^{-1}(\pi(\xi(x^{-1},x))\pi(f(x^{-1}))U(x)^*\Omega,\Omega)\cr
&=&\int_X\hbox{ d}x\,(\pi(f(x))U(x)\Omega,\Omega)
\end{eqnarray}

A representation $\overline\pi$ of ${\cal A}\times_\xi X$
is now defined by extension of
\begin{equation}
\overline\pi(f)=\int_X\hbox{ d}x\,\pi(f(x))U(x),
\qquad f\in {\cal L}_1(X,{\cal A})
\end{equation}
It is obvious that this representation is equivalent with the
GNS-rep\-resen\-tation of $\overline\omega$. Conversely, if
$\overline\omega$ is any state of ${\cal A}\times_\xi X$ then a
state $\omega$ of the covariance system $({\cal A},X,\sigma)$ can
be defined by (9) using the GNS-representation of $\overline\omega$
(see the remarks after Theorem 2 of \cite{NJ99}). This representation
is then equivalent with the GNS-representation of $\omega$.

\section{Nonrelativistic particle}

This section is devoted to projective representations
occurring in quantum mechanics of a nonrelativistic particle
and serves as an example of our approach. It relies heavily on
older work by Levy-Leblond\cite{LLB63}, work which has been extended by
Hagen\cite{HR70,HR71}.

The algebra $\cal A$ is the algebra ${\cal C}_0(\Ro^n)$
of classical functions of position. The appropriate
group of symmetry transformations is the Galilei group.
However, it is easier to start with the subgroup $\Ro^n,+$
of shifts. This yields already the standard representation of
quantum mechanics. Later on rotations and transformations
to a moving frame are added to discuss spin and mass of the particle.

\subsection{Standard representation}
\label{standrepr}

First assume $X=\Ro^n,+$.
The shifts act as automorphisms of $\cal A$ by
\begin{equation}
(\sigma_qf)(q')=f(q'-q),
\qquad f\in {\cal A}, q,q'\in X
\end{equation}
A *-representation $\pi$ of $\cal A$ as bounded operators of
${\cal H}={\cal L}_2(\Ro^n)$ is defined by
\begin{equation}
\pi(f)\psi(q)=f(q)\psi(q),
\qquad f\in {\cal A}, q\in\Ro^n
\end{equation}
A unitary representation $U$ of $X$ is defined by
\begin{equation}
U(q)\psi(q')=\psi(q'-q),
\qquad q,q'\in\Ro^n
\end{equation}
One verifies immediately that $({\cal H},\pi,U)$
is a representation of $({\cal A},X,\sigma)$.
In particular, any normalized element $\psi$ of $\cal H$ defines a
state of $({\cal A},X,\sigma)$. The multiplier $\xi$ associated with
such a state is identically equal to 1.

Using Stone's theorem the shift operators can be written as
\begin{equation}
U(q)=\exp(-(i/\hbar)\sum_{j=1}^nq_jP_j)
\end{equation}
with $P_j$ the momentum operators and with $\hbar$ equal
to Planck's constant divided by $2\pi$.
A short calculation gives the canonical commutation relations
\begin{equation}
\left[Q_j,P_k\right]_-=\delta_{j,k}i\hbar
\end{equation}
with $Q_j$ the multiplication operators
\begin{equation}
Q_j\psi(q)=q_j\psi(q)
\end{equation}
defined on a suitable domain.

\subsection{Rotation symmetry}

We extend the symmetry group $X$ now to shifts and rotations.
We first discuss a particle without spin.

The group law is
\begin{equation}
(q',\Lambda')(q,\Lambda)=(q'+\Lambda' q,\Lambda'\Lambda),
\qquad q,q'\in\Ro^n,\Lambda,\Lambda'\in {\rm SO}(n)
\end{equation}
The representation $\sigma$ is given by
\begin{equation}
(\sigma_{q,\Lambda}f)(q')=f(\Lambda^{-1}(q'-q)),
\qquad f\in{\cal A},q,q'\in \Ro^n,\Lambda\in {\rm SO}(n)
\end{equation}
The *-representation $({\cal H},\pi)$ is the standard representation
of the previous section. The unitary representation $U$ of $X$
is given by
\begin{equation}
(U(q,\Lambda)\psi)(q')=\psi(\Lambda^{-1}(q'-q)),
\qquad\psi\in{\cal L}_2(\Ro^n),q,q'\in \Ro^n,\Lambda\in {\rm SO}(n)
\label{rotshift}
\end{equation}
Then $({\cal H},\pi,U)$
is a representation of $({\cal A},X,\sigma)$.
In particular, any normalized element $\psi$ of $\cal H$ defines a
state of $({\cal A},X,\sigma)$. The multiplier $\xi$ associated with
such a state is identically equal to 1.

\subsection{Spin}

Fix $n=3$. The covariance system $({\cal A},X,\sigma)$ of the previous 
section has also states whose associated multiplier $\xi$ is nontrivial. The 
representation induced by these states is the so-called spinor representation. 
The reason of their existence is that the group SO$(3)$ is double connected, 
with covering group SU$(2)$.

Given $q\in\Ro^3$, construct the matrix
\begin{equation}
M(q)=\sum_{j=1}^3q_j\sigma_j
\end{equation}
with $\sigma_1,\sigma_2,\sigma_3$ the three Pauli matrices.
The matrix $M(q)$ transforms under an element $u$ of SU$(2)$
into the matrix $M(q')=uM(q)u^*$. It is easy to show that the
transformation $q\rightarrow q'$ is a rotation, i.e.~there
exists $\Xi(u)\in{\rm SO}(3)$ for which $q'=\Xi(u) q$.
Note that $\Xi$ is a homomorphism of SU$(2)$ onto SO$(3)$.
Let $\Lambda\rightarrow v(\Lambda)$ be an inverse of
$u\rightarrow \Xi(u)$ in the sense that $\Xi(v(\Lambda))=\Lambda$
for all $\Lambda$ and $v(\Xi(u))=u$ for all $u$ in a
neighborhood of the identity matrix. Note that the map $v$ cannot
be continuous.  A cocycle $\xi$ of SO$(3)$ is defined by
\begin{equation}
v(\Lambda)v(\Lambda')=\xi(\Lambda,\Lambda')v(\Lambda\Lambda')
\end{equation}
From
\begin{equation}
\Xi(v(\Lambda)v(\Lambda'))=\Lambda\Lambda'
\end{equation}
and
\begin{equation}
\Xi(\xi(\Lambda,\Lambda')v(\Lambda\Lambda'))=
\Xi(\xi(\Lambda,\Lambda'))\Lambda\Lambda'
\end{equation}
follows that
$\Xi(\xi(\Lambda,\Lambda'))=\Io$. This implies $\xi(\Lambda,\Lambda')=\pm 1$.

Consider the Hilbert space ${\cal H}={\cal L}_2(\Ro^n)\oplus{\cal L}_2(\Ro^n)$.
An element of this space is denoted
$\displaystyle
\left(\begin{array}{c}
\psi_1\\ \psi_2
\end{array}\right)
$.
It is normalized if $||\psi_1||^2+||\psi_2||^2=1$.
A *-representation $\pi$ of $\cal A$ is defined by
\begin{equation}
\pi(f)\left(\begin{array}{c}
\psi_1\\ \psi_2
\end{array}\right)(q)=
f(q)\left(\begin{array}{c}
\psi_1\\ \psi_2
\end{array}\right)(q)
\end{equation}
A projective representation $U$ of $X$ is defined by
\begin{equation}
U(q,\Lambda)\left(\begin{array}{c}
\psi_1\\ \psi_2
\end{array}\right)(q')=
v(\Lambda)\left(\begin{array}{c}
\psi_1\\ \psi_2
\end{array}\right)(\Lambda^{-1}(q'-q))
\end{equation}
Indeed, one verifies that
\begin{eqnarray}
& &U(q',\Lambda')U(q,\Lambda)\left(\begin{array}{c}
\psi_1\\ \psi_2
\end{array}\right)(q'')\cr
&=&v(\Lambda')v(\Lambda)\left(\begin{array}{c}
\psi_1\\ \psi_2
\end{array}\right)
(\Lambda^{-1}({\Lambda'}^{-1})(q''-q')-q))\cr
&=&\xi(\Lambda',\Lambda)v(\Lambda'\Lambda)
\left(\begin{array}{c}
\psi_1\\ \psi_2
\end{array}\right)
((\Lambda'\Lambda)^{-1}(q''-q'-\Lambda' q)\cr
&=&\xi(\Lambda',\Lambda)U(q'+\Lambda' q,\Lambda'\Lambda)
\left(\begin{array}{c}
\psi_1\\ \psi_2
\end{array}\right)(q'')
\end{eqnarray}
Because (\ref{covariant}) is satisfied it follows from prop.~\ref{projprop}
that $({\cal H},\pi,U)$ is a representation of $({\cal A},X,\sigma)$.

Each normalized element $\displaystyle \left(\begin{array}{c}
\psi_1\\ \psi_2
\end{array}\right)$ of 
$\cal H$ determines a state $\omega$ of $({\cal A},X,\sigma)$.
A tedious calculation yields
\begin{eqnarray}
\omega_{q,\Lambda;q',\Lambda'}(f) &=&\int_{\Ro^3}\hbox{ d}q''\,f(q'')\cr
& &\times\left(\begin{array}{c}
\psi_1\\ \psi_2
\end{array}\right)(\Lambda'q''+q')
\cdot v(\Lambda')v(\Lambda)^*\left(\begin{array}{c}
\psi_1\\ \psi_2
\end{array}\right)(\Lambda q''+q)
\end{eqnarray}
In particular,
\begin{eqnarray}
\omega_{q,\Lambda;q,\Lambda}(f) &=&\int_{\Ro^3}\hbox{ d}q''\,f(q'')
(|\psi_1|^2+|\psi_2|^2)(\Lambda q''+q)
\end{eqnarray}
These diagonal elements of $\omega$ do not reveal whether the state has spin or not.
Only correlation effects visible in the off-diagonal elements
can reveal the presence of mechanical spin.
Consider e.g.~a rotation 
by $\pi$ around the $z$-axis (denoted $\Lambda$). Then one has
$v(\Lambda)=\sigma_z$ so that
\begin{equation}
\omega_{q,\Lambda;0,\Io}(f)=
\int_{\Ro^3}\hbox{ d}q''\,f(q'')
(\overline\psi_1(q'')\psi_1(\Lambda q''+q)-
\overline\psi_2(q'')\psi_2(\Lambda q''+q))
\end{equation}
The minus sign is a consequence of spin and has the
effect that the particle observed in a frame
rotated by 180 degrees does not look like expected.

\subsection{Mass}

From now on $X$ is the full Galilei group. Its composition law is
\begin{equation}
(q',\Lambda',t',v')(q,\Lambda,t,v)=
(q'+\Lambda' q+tv',\Lambda'\Lambda,t'+t,v'+\Lambda' v)
\end{equation}
As before $q\in\Ro^n,+$ is a shift operation, and $\Lambda\in{\rm SO}(3)$
is a rotation. The remaining parameters describe a
transformation to a moving frame with velocity $v\in\Ro^n$ and time $t\in\Ro$.

For simplicity we consider a particle without spin. Then the relevant *-
representation $\pi$ of $\cal A$ is the standard representation of section 
\ref{standrepr}. The Galilei group has 
nontrivial cocycles $\xi_\kappa$ parameterized by some parameter $\kappa\not=0$
\cite{BV54,LLB63}.
A possible definition of $\xi_\kappa$ is
\begin{equation}
\xi_\kappa(q',\Lambda',t',v';q,\Lambda,t,v)=
\exp\left(-i\kappa\left[
 {1\over 2}t'|v|^2 + t' v'\cdot\Lambda'v
- v'\cdot\Lambda'(q-tv)
\right]\right)
\end{equation}
It appears in the following projective representation of $X$
\begin{equation}
U(q,\Lambda,t,v)
=U(q-tv,\Lambda)\exp(i\kappa v\cdot \Lambda Q)\exp(i\hbar^{-1}tH)
\end{equation}
where $H$ is the hamiltonian of the free particle
\begin{equation}
H={1\over 2m}P^2
\end{equation}
the mass $m$ is related to $\kappa$ by $m=\hbar\kappa$,
$P$ and $Q$ are the position and momentum operators
(see section \ref{standrepr}), and $U(q,\Lambda)$ is the unitary
representation defined by (\ref{rotshift}).
Indeed, a tedious calculation shows that $U$,
defined in this way, satisfies (\ref{projrepr})
with $\xi=\xi_\kappa$. Now (\ref{covariant}) can
be taken as the definition of the representation
$\sigma$ of $X$ as automorphisms of $\cal A$.
By proposition \ref{projprop} the triple
$({\cal H},\pi,U)$ is a representation of the
covariance system $({\cal A},X,\sigma)$.
Any normalized element $\psi$ of $\cal H$ determines
a state $\omega$ of $({\cal A},X,\sigma)$.
The associated $C^*$-multiplier is $\xi_\kappa$.
The state $\omega$ describes a free particle with mass $m=\hbar\kappa$.

\section{Noncanonical commutation relations}

In this section we construct a variant of the model introduced by 
Doplicher et al \cite{DFR94,DFR95}. It describes a single quantum particle
in spacetime. We do not follow the usual notational conventions of relativity
theory. In particular, the inner product $x\cdot y$ is that of $\Ro^4$
and not that of Minkowski space. The reason for doing so is that we use
simultaneously two different metric tensors, denoted $g$ resp.~$\gamma$.

\subsection{Model}

Let us make the (quite unusual) {\sl ansatz}
that a relativistic particle is characterized by
two elements $e$ and $m$ of $\Ro^3$ satisfying $|e|^2=|m|^2$
and $e\cdot m=\pm 1$. Let $\Sigma\subset\Ro^6$ denote the
space of these pairs $(e,m)$.
It is by assumption the classical configuration space of the particle.
See \cite{DFR94,DFR95} for a motivation of this particular
choice. Note that $\Sigma$ is a locally compact manifold in $\Ro^6$.
The algebra $\cal A$ of classical observables equals the $C^*$-algebra
${\cal C}_0(\Sigma)$ of continuous
complex functions of $\Sigma$ vanishing at infinity.
Consider as symmetry group $X$ of $\Sigma$ the group $\Ro^4\times\Ro^4,+$
acting in a trivial way, i.e.~$\sigma_{k,q}f=f$
for all $k,q\in\Ro^4$ and $f\in{\cal C}_0(\Sigma)$. In this
way one obtains a covariance system $({\cal A},X,\sigma)$.

The metric tensor of Minkowski space is denoted $g$. It
is the diagonal matrix $[1,-1,-1,-1]$.
The geometry of the space $\Sigma$ is described by a metric tensor
$\gamma(e,m)$ which is an invertible 4-by-4 matrix with
real coefficients depending continuously on $(e,m)$.
A possible choice is $\gamma=g$.

Later on the group $X$ will be extended with the Lorentz group.
The action of the Lorentz group on functions of $\Sigma$ will be
nontrivial. Since integration over $\Sigma$ should be compatible
with this action we introduce it already now.

For each $(e,m)\in\Sigma$ form an antisymmetric matrix $\epsilon(e,m)$ by
\begin {equation}
\epsilon(e,m)=\left(\begin{array}{cccc}
0    &e_1  &e_2  &e_3 \cr
-e_1 &0    &m_3  &-m_2\cr
-e_2 &-m_3 &0    &m_1 \cr
-e_3 &m_2  &-m_1 &0\cr
\end{array}\right)
\end {equation}
The matrix $\epsilon^{-1}\gamma$ will appear often.
It transforms positions $q$ into wave vectors $k$.
Let us introduce the notation
\begin{equation}
\eta(e,m)=(e\cdot m)\epsilon^{-1}(e,m)\gamma(e,m)
\end{equation}
The factor $(e\cdot m)$, which equals $\pm 1$, has
been included for physical reasons (one expects that
the wave vector $k$ will transform into $gk$ under
time reversal, while $q$ transforms into $-gq$).

A representation $\sigma$ of the Lorentz group as automorphisms
of ${\cal C}_0(\Sigma)$ is defined by
\begin{equation}
\sigma_\Lambda f(e,m)=f(e',m')
\qquad\hbox{ with }
\epsilon(e',m')=\tilde \Lambda\epsilon(e,m)\Lambda
\label{stsigma}
\end{equation}
The Haar measure of the Lorentz group induces a measure $\rho_0$ of
$\Sigma$ because each point of $\Sigma$ can be reached by
transforming a fixed point, say $e=m=(1,0,0)$.

\subsection{Quasifree states}

Let $T(e,m)$ be a positive matrix, continuously depending on
$(e,m)$, large enough so that
$T(e,m)+(i/2)(e\cdot m)\epsilon(e,m)$ is a positive definite matrix
(an appropriate 'covariant' choice of $T$ is discussed later on).
Let $w$ be a strictly positive function with integral 1 and let
\begin{equation}
\hbox{d}\rho(e,m)=w(e,m)\hbox{ d}\rho_0(e,m)
\end{equation}
The probability measure $\rho$ together with the choice of
$T$ determines a state $\omega$ of $({\cal A},X,\sigma)$ by
\begin{eqnarray}
& &\omega_{k,q;k',q'}(f)\cr
&=&\int_\Sigma\hbox{ d}\rho(e,m)\,
f(e,m)\,\exp\left(
{i\over 2}(e\cdot m)(k+\eta q)\cdot \epsilon(k'+\eta q')
\right)\cr
&\times&\exp\left(-{1\over 2}(k-k'+\eta (q-q'))\cdot T
(k-k'+\eta (q-q'))
\right)
\label{quasifree}
\end{eqnarray}
for all $k,k',q,q'\in\Ro^4$
The $C^*$-multiplier $\xi$ associated with $\omega$ is given by
\begin{eqnarray}
\xi(k,q;k',q')
&=&\exp\left(
{i\over 2} (e.m)(k+\eta q)\cdot\epsilon(k'+\eta q')
\right)
\end{eqnarray}
Note that in (\ref{quasifree}) $k,k'$ has the interpretation of a shift in
wave vector, $q,q'$ that of a shift in position of the particle.

\subsection{Position and momentum operators}

Let $({\cal H},\pi,U,\Omega)$ be the GNS-representation of $({\cal A},X,\sigma)$
induced by $\omega$. From the continuity of (\ref{quasifree}) in $k,k',q,q'$
follows, using Stone's theorem, that $U$ has self-adjoint generators.
Because $\xi(k,q;k',q')$ commutes with $U(k'',q'')$ one can write
\begin{equation}
U(k,q)=\exp\left(-ik\cdot \gamma Q+i\gamma q\cdot K\right)
\label{ugenqst}
\end{equation}
In the context of the GNS-construction the operators $Q_\mu$ and $K_\mu$
can be written formally as a sum of a multiplication operator and
a differential operator, acting on functions $\psi(k,q)$ with value in
${\cal C}_0(\Sigma)$,
From the definition (see the proof of the GNS-construction)
follows
\begin{equation}
U(k,q)\psi(k',q')=\psi(k+k',q+q')\xi(k',q';k,q)
\label{udefqst}
\end{equation}
Comparing expansions of (\ref{ugenqst}) and (\ref{udefqst}) one obtains
\begin{equation}
Q_\mu
=i\sum_{\nu}\gamma^{-1}_{\mu,\nu}{\partial\,\over\partial k_{\nu}}
+{1\over 2}q_\mu
+{1\over 2}(\eta^{-1} k)_\mu
\end{equation}
and
\begin{equation}
K_\mu
=-i\sum_{\nu}\gamma^{-1}_{\nu,\mu}{\partial\,\over\partial q_{\nu}}
+{1\over 2}k_\mu
+{1\over 2}(\eta q)_\mu
\end{equation}
Using these explicit expressions it is easy to calculate the
following commutation relations.
\begin{eqnarray}
\bigg[ Q_\mu,Q_\nu\bigg]
&=&-i(e\cdot m)(\gamma^{-1}\epsilon\tilde\gamma^{-1})_{\mu,\nu}\cr
\bigg[ K_\mu,K_\nu\bigg]
&=&i(e\cdot m)\epsilon^{-1}_{\mu,\nu}\cr
\bigg[ K_\mu,Q_\nu\bigg]
&=&-i\gamma^{-1}_{\nu,\mu}
\end{eqnarray}
Note that the r.h.s.~of these expressions is always an operator commuting
with all $Q_\mu$ and $K_\mu$.
Commutation relations of this kind have been proposed by
Doplicher et al \cite{DFR94,DFR95} as a simplified model for quantum spacetime.

It is now very easy to calculate physically relevant quantities
in the state $\omega$. E.g, one finds $(Q_\mu\Omega,\Omega)=0$
and
\begin{eqnarray}
(Q_\nu Q_\mu\Omega,\Omega)
&=&
\int_\Sigma\hbox{ d}\rho(e,m)
\left[\gamma^{-1}
\left(T(e,m)+{i\over 2}(e\cdot m)\epsilon(e,m)\right)
\gamma^{-1}\right]_{\mu,\nu}
\label{posuncert}
\end{eqnarray}
i.e., the state $\omega$ describes a particle with expected position at the
origin of spacetime and with variance in position determined by $T\pm (i/2)\epsilon$.

\subsection{Lorentz transformations}

In what follows we assume that
\begin{equation}
T(e',m')=\tilde \Lambda T(e,m)\Lambda
\label{Tcov}
\end{equation}
with $(e',m')$ related to $(e,m)$ by (\ref{stsigma}),
i.e.~ $T$ transforms like $\epsilon$ under Lorentz transformations.
Such an assumption is possible. Indeed, let
$C$ be any positive matrix. Fix a special point $(e_0,m_0)$ in $\Sigma$,
e.g.~the point with $e=m=(1,0,0)$. Use it to define the matrix $\epsilon_0$
by $\epsilon_0=\epsilon(e_0,m_0)$.
Now assume that $\tilde\Lambda\epsilon_0\Lambda=\epsilon_0$ implies that
$\tilde\Lambda C\Lambda = C$. Then $T(e,m)$ is defined by
\begin{equation}
T(e,m)=\tilde\Lambda C \Lambda
\qquad\hbox{ whenever }\epsilon(e,m)=\tilde\Lambda\epsilon_0\Lambda
\end{equation}
Obviously, $T$ transforms in the same way as $\epsilon$.
From
\begin{equation}
k\cdot T(e,m)k'=\Lambda k \cdot C \Lambda k'
\end{equation}
follows that $T(e,m)$ is positive.
In fact we need $T(e,m)+(i/2)(e\cdot m)\epsilon(e,m)\ge 0$.
This is satisfied if $C \pm (i/2)\epsilon_0\ge 0$. E.g., $C=(1/2){\Io}$
is a good choice.

If (\ref{Tcov}) is satisfied then a unitary representation $R$ of the
proper Lorentz group is defined by
\begin{equation}
R(\Lambda)\psi(k,q)(e,m)=\sqrt{{w(e',m')\over w(e,m)}}
\psi(\Lambda^{-1} k,\Lambda^{-1}q)(e',m')
\end{equation}
with $(e',m')$ as in (\ref{stsigma}).
One calculates that
\begin{equation}
R(\Lambda)^*\psi(k,q)(e',m')=
\sqrt{{w(e,m)\over w(e',m')}}\psi(\Lambda k,\Lambda q)(e,m)
\end{equation}
It is now straightforward to verify that $R(\Lambda)$ is unitary.
One verifies that
\begin{equation}
\pi(\sigma_\Lambda f)=R(\Lambda)\pi(f)R(\Lambda)^*
\end{equation}
With the assumption that
\begin{equation}
\gamma(e',m')=\tilde\Lambda\gamma(e,m)\Lambda
\end{equation}
i.e.~also $\gamma$ transforms in the same way as $\epsilon$,
one obtains
\begin{equation}
U(\Lambda k,\Lambda q)=R(\Lambda)U(k,q)R(\Lambda)^*
\end{equation}

Extension of the symmetry group $X$ to include the proper Lorentz group
is straightforward. E.g., let
\begin{equation}
\omega_{k,q,\Lambda;k',q',\Lambda'}(f)=
(\pi(f)U(k,q)^*R(\Lambda)^*\Omega,U(k',q')^*R(\Lambda')^*\Omega)
\end{equation}

\section{Discussion}

We have introduced the notion of a (mathematical) state of a covariance system 
$({\cal A},X,\sigma)$. It generalizes the notion of a covariant state of the 
$C^*$- algebra $\cal A$ by allowing that in the Hilbert space representation the 
action $\sigma$ of the group $X$ is implemented as a projective representation 
and by allowing that the cocycle associated with this representation is operator 
valued (i.e., it is a $C^*$-multiplier). For these generalized covariant states 
we prove a GNS-theorem. It can be used as an alternative for working with 
crossed products of $\cal A$ and $X$. In fact, the proof captures all essential 
elements of the proof of the existence of the crossed product algebra. The shift 
in emphasis from crossed product algebra to states of a covariance system has 
many advantages. In particular, the problem that the crossed product algebra 
depends on the $C^*$-multiplier is circumvented. E.g., mass and spin of a 
quantum particle are properties of states of a covariance system, involving 
different $C^*$-multipliers. In the more traditional approach particles with
or without spin are described by states on different $C^*$-algebras.

A limitation of the present work is that representations involving anti-unitary 
operators are not included. Such representations are essential for physical 
applications. Wigner \cite {WEP31} showed that all symmetry elements appearing 
in quantum mechanics must be implemented either by unitary or anti-unitary 
operators of the Hilbert space of wave functions. These anti-unitary operators 
appear only in case of discrete symmetries. Indeed, one can always assume that 
the neutral element of the symmetry group is implemented as the identity 
operator. Then any element which is continuously connected with the neutral 
element must also be implemented as a unitary operator. We think that the
omission of anti-unitary representations can be handled easily in most situations.
It is clear from the 
examples that the physically relevant representation of the covariance system is 
usually already fixed by considering a subgroup of symmetries (e.g., the 
standard representation of quantum mechanics is already obtained by considering 
only the subgroup of spatial shifts). In such cases the anti-unitary 
implementation of discrete symmetry elements can be added 'by hand', 
i.e.~without relying on the GNS-theorem.A good example\cite {SJ51} is 
time reversal symmetry, denoted $\theta$. When added to the Galilei group it 
anti-commutes with velocity $v$ and time $t$, and commutes with position $q$ 
and rotation $\Lambda$. It is implemented as the anti-unitary operator which 
maps each wave function onto its complex conjugate. In the relativistic example 
$\theta=-g$ is an element of the full Poincar\'e group. Its implementation is
\begin{equation}
R(\theta)\psi(k,q)(e,m)
=\sqrt{{w(-e,m)\over w(e,m)}}\overline{\psi(g k,\theta q)(-e,m)}
\end{equation}

The section on non-relativistic quantum mechanics picks up old ideas about the 
role of covariance systems in quantum mechanics and illustrates our approach to 
quantum mechanics and quantum field theory. It does not involve a quantization 
step referring to classical mechanics. It starts from a $C^*$-algebra $\cal A$ 
of 'classical' functions which are accessible for experimental measurement. A 
group $X$ of symmetries, acting as automorphisms of $\cal A$, is considered. 
Together they form a covariance system. E.g., if $\cal A$ is a $C^*$-algebra of 
functions of position in $\Ro^3$ and $X$ is the Galilei group then the standard 
representation of quantum mechanics is a (projective) representation of the 
covariance system $({\cal A}, X,\sigma)$. 

The third section about the model of quantum spacetime has been added to give an 
example of projective representations involving operator valued cocycles 
i.e.~non-trivial $C^*$-multipliers. The immediate effect of these are
non-canonical commutation relations. Another aspect, well illustrated by this 
example, is how the GNS-construction can be used to build physically interesting 
representations starting from an explicit expression for a state of the 
covariance system. It is quite clear that our treatment of this model is far 
from complete. A full study is out of scope of the present paper
and will be reported elsewhere.

\begin{acknowledgement}
This work benifited from financial support by the Bilateral Scientific
and Technological Cooperation Programme between Poland and Flanders
and by the Research Council of the University of Antwerp.
\end{acknowledgement}

\vfill\eject

\end{document}